\documentclass[conference]{IEEEtran}
\IEEEoverridecommandlockouts
\ifCLASSOPTIONcompsoc
\usepackage[caption=false,font=normalsize,labelfon
t=sf,textfont=sf]{subfig}
\else
\usepackage[caption=false,font=footnotesize]{subfi
g}
\fi
\usepackage{cite}
\usepackage{amsmath,amssymb,amsfonts}
\usepackage{algorithmic}
\usepackage{graphicx}
\usepackage{textcomp}
\usepackage{xcolor}
\usepackage{graphicx}
\usepackage{float}
\usepackage{caption}
\usepackage{lipsum}
\usepackage{physics}

\def\BibTeX{{\rm B\kern-.05em{\sc i\kern-.025em b}\kern-.08em
    T\kern-.1667em\lower.7ex\hbox{E}\kern-.125emX}}
\begin{document}

\title{Variational Quantum Algorithms for Chemical Simulation and Drug Discovery}

\makeatletter
\newcommand{\linebreakand}{%
  \end{@IEEEauthorhalign}
  \hfill\mbox{}\par
  \mbox{}\hfill\begin{@IEEEauthorhalign}
}
\makeatother

\author{\IEEEauthorblockN{Hasan Mustafa}
\IEEEauthorblockA{\textit{AI@Scale, Fractal Analytics}\\
Mumbai, India \\
hasan.mustafa@fractal.ai}
\and
\IEEEauthorblockN{Sai Nandan Morapakula}
\IEEEauthorblockA{\textit{AI@Scale, Fractal Analytics}\\
Bangalore, India\\
sai.nandan@fractal.ai}
\and
\IEEEauthorblockN{Prateek Jain}
\IEEEauthorblockA{\textit{AI@Scale, Fractal Analytics}\\
Gurugram, India \\
prateek.jain@fractal.ai}
\linebreakand
\IEEEauthorblockN{Srinjoy Ganguly}
\IEEEauthorblockA{\textit{AI@Scale, Fractal Analytics}\\
Gurugram, India\\
srinjoy.ganguly@fractal.ai}
}

\maketitle

\begin{abstract}

Quantum computing has gained a lot of attention recently, and scientists have seen potential applications in this field using quantum computing for Cryptography and Communication to Machine Learning and Healthcare.
Protein folding has been one of the most interesting areas to study, and it is also one of the biggest problems of biochemistry. Each protein folds distinctively, and the difficulty of finding its stable shape rapidly increases with an increase in the number of amino acids in the chain. A moderate protein has about 100 amino acids, and the number of combinations one needs to verify to find the stable structure is enormous.  
At some point, the number of these combinations will be so vast that classical computers cannot even attempt to solve them. In this paper, we examine how this problem can be solved with the help of quantum computing using two different algorithms, Variational Quantum Eigensolver (VQE) and Quantum Approximate Optimization Algorithm (QAOA), using Qiskit Nature. We compare the results of different quantum hardware and simulators and check how error mitigation affects the performance. Further, we make comparisons with SoTA algorithms and evaluate the reliability of the method.
\end{abstract}


\section{Introduction}
Proteins are chains of smaller molecules called amino acids. Amino acids bond with each other by making a so-called Peptide bond. The median length of proteins in terms of the number of amino acids is 375 for Humans. These incredibly long protein chains fold and form complex 3-D structures that are stable. Finding this stable 3-D shape of a particular protein given the amino acid sequence is known as the \textbf{protein folding problem.}
Understanding how proteins fold is essential as the function of a protein is correlated with its shape. The 3-D structure of the protein is entirely determined by the amino acid sequence. Thus, mutations which can lead to a change in the amino acid sequence of a protein cause the protein to fold differently and render it incapable of performing the desired function. The effects of a misfolded protein can range from minor diseases to severe illnesses. 
Some of the diseases which are caused by improperly folded proteins are,
Type 2 Diabetes, Alzheimer's, Parkinson's, Amyotrophic Lateral Sclerosis(ALS). The amino acid sequence considered resembles a protein that interacts with Alzheimer's disease-causing enzyme. 

Drug development involves simulating the behaviour of the proteins and their interaction with drug molecules.
Being able to simulate the 3-D structure of protein reliably would considerably speed up the drug discovery process, which currently relies on experimental methods that are expensive and time-consuming. The protein folding problem is computationally very expensive and scales exponentially in the number of amino acids. This complexity has led to the development of new techniques such as those using machine learning. The most successful of them being AlphaFold, which outperformed other techniques in benchmarking competition \cite{c1}.

Through the paper, we provide a comprehensive analysis of the VQE in contrast to the SoTA AlphaFold model. This is one of the first of its kind benchmarking result for a particular application on different hardware. We also identify the caveats in the existing approach and quantify them. 

In this paper, we examine the application of Variational Quantum Algorithms to the protein folding problem. We use the protein folding module developed by Qiskit \cite{c2} to model the problem. Variational algorithms are well suited for the Noisy intermediate-scale quantum(NISQ) era, which is characterized by noisy hardware capable of running low-depth circuits. 
% These numbers increase exponentially with the length of the amino acid chain. A typical protein has about 10\textsuperscript{100} ways in which it can fold, and finding the right one from all of these, checking one at a time is a tedious task. The exponential growth of the foldings with the chain length makes the job difficult for classical computers. Recently, scientists found out that proteins fold to that particular shape where it has the lowest energy state. We take this to our advantage and use quantum resource-efficient algorithms such as VQE and QAOA which are specifically designed to find the ground state of a system to do the work. The main goal is to determine the minimum energy conformation of the protein with various error mitigation techniques. 

The paper is organized as follows. Section \ref{lit_rev} outlines the protein folding model and the quantum algorithms we use. In section \ref{meth_set}
we describe the specific setting we have used such as the optimizers and ansatz. We describe all the hardware and simulator results of various types of quantum computers and compare them in section \ref{result}. Section \ref{discussion} comments on the results. Finally we give additional comments and summary in Section \ref{conclusion}.%\ref{future_research} comments on the future scope of research.

\section{Literature Review} \label{lit_rev}

% The protein folding model we consider in this paper is a lattice-based model developed by . By using a tetrahedral lattice the search space for conformation is drastically reduced by eliminating off-grid positions. 

\subsection{Protein Folding Problem}
The protein folding problem seeks to answer the question of how a protein's 3-D structure can be determined from the amino acid sequence \cite{c3}. This can be modelled as a search problem in the space of all possible conformations. The protein folding problem is characterized by an infinitely large search space. The solution to this problem represents the functional structure of the protein that matches the experimentally determined structure. The number of possible conformers can go as high as 100\textsuperscript{300}. The enormity of the search space makes it NP-Hard.
It is well established that protein folds into a thermodynamically stable conformation \cite{c3}. Thus, the goal of the search problem can be equivalently set to finding the protein structure with the minimum energy\cite{c4}.

\subsection{Protein Folding Model}
The peptide is abstracted as a sequence of beads on a tetrahedral lattice. Each point in a tetrahedral lattice is connected to 4 other points; thus the chain can "turn" in one of the four directions when the following amino acid is added. The direction in which the chain turns with the addition of each amino acid from the sequence is used to describe a protein conformation. The amino acid chain is encoded by assigning 4 qubits per bead, each qubit representing one of the four possible turns. Without loss of generality, the first two turns can be chosen arbitrarily. The remaining chain can then be encoded into 4(N-3) qubits. The representation was further condensed to reduce the number of qubits used to 2(N-3).
The interaction between the amino acids is calculated based on inter-residue contact energies evaluated by Miyazawa-Jernigan \cite{c5}. Interaction qubits are separately used to encode the interaction energy.
Three penalty terms, penalty chiral, penalty back, penalty 1, are also introduced to prevent the chain from folding back to itself, impose chirality constraints, and penalize local overlap between beads within a nearest neighbour contact.
A Hamiltonian encoding the interaction energies and the penalties for which the ground state is to be determined is generated.
\[H(\mathbf{q}) = H_{gc}(\mathbf{q}_{cf}) + H_{ch}(\mathbf{q}_{cf}) + H_{in}(\mathbf{q}_{cf}, \mathbf{q}_{in})\]
\begin{itemize}
    \item \(H_{gc}\) is the geometrical constraint term (governing the growth of the primary sequence of amino acids without bifurcations)
    \item \(H_{ch}\) is the chirality constraint (enforcing the right stereochemistry for the system)
    \item \(H_{in}\) is the interaction energy terms of the system. In our case we consider only nearest neighbor interactions.
\end{itemize}

Two of the most common variational algorithms that achieve the objective are VQE and QAOA. A subsequent modification of these algorithms is to use CVaR \cite{c6} as the objective function instead of simply the expectation value.

\subsection{VQE}
Variational Quantum Eigensolver(VQE) is a hybrid classical-quantum algorithm that finds the eigenvalue of the ground state of a Hamiltonian. VQEs are based on the variational principle. VQEs involve estimating the expectation value of the Hamiltonian for a parametric ansatz state and then minimizing the expectation value by using a classical optimizer to tune the parameters. The ansatz state might be chosen based on the physical properties of the system
\cite{c7}.  Quantum Phase Estimation(QPE) algorithm can also be used for the same purpose, but its implementation is not feasible in the NISQ Era and is thus not considered in this paper.
\begin{figure}[htbp]
 \centering
 \includegraphics[scale=.24]{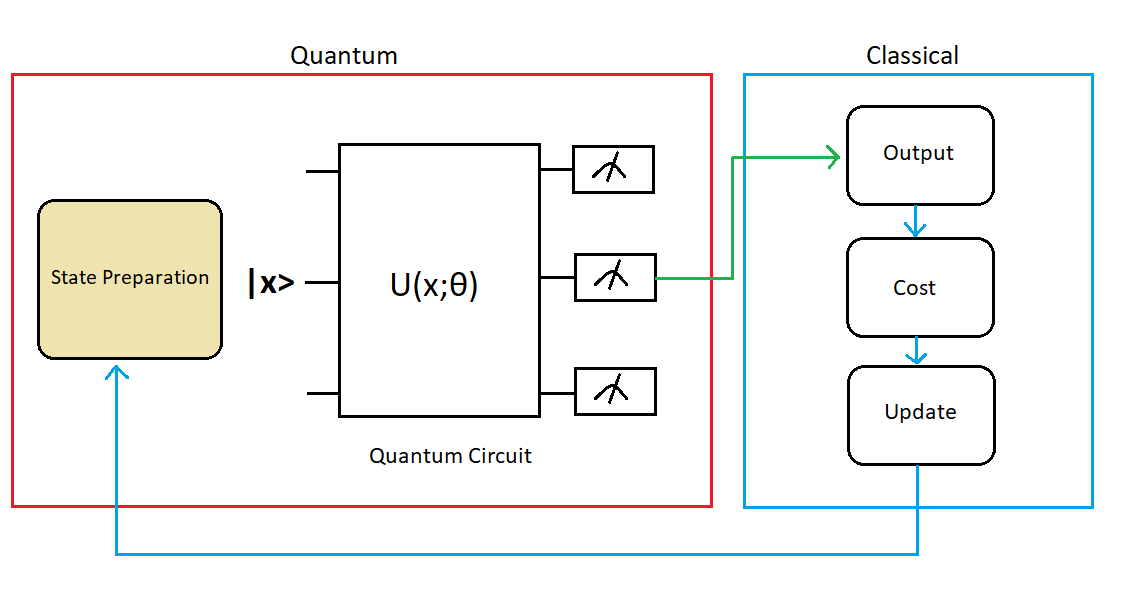}
 %\setcaptioncitation{\textit{Source:} Qiskit Nature}
  %remove the above line? and add to caption?
 \caption{Working of VQE}
 \label{fig1}
\end{figure}

Figure \ref{fig1} illustrates the workflow of VQE as split between the classical and quantum computers. We use classical computers/optimizers to adjust and tune the parameters. 
VQE has three main components, Ansatz, Hamiltonian, and Measurement. An ansatz is the parametrized quantum circuit that generates the quantum state we need in order to apply the variational algorithm. It should be noted that an ansatz is not fixed; the design of the ansatz depends on the problem and hardware; the better the range and its shallowness the better is the ansatz. Hamiltonian, denoted as a matrix or a sum of Pauli operators. 
\subsection{QAOA}
Quantum Approximate optimization algorithm(QAOA) was proposed to solve combinatorial optimizations. The optimization problem can be remodelled into an equivalent problem of finding the ground state of the corresponding Hamiltonian. The variational form chosen for QAOA is\cite{c8}
\begin{equation}
    U(C,\gamma_1)U(M,\beta_1)..p times...U(C,\gamma_p)U(M,\beta_p)\ket{\Psi}
\end{equation}
\begin{equation}
U(C,\gamma)=e^{-iH_C\gamma}
\end{equation}
\begin{equation}
    U(M,\beta)=e^{-iH_B\beta}
\end{equation}
Where \(H_C\) and \(H_{B}\) are the cost and mixing Hamiltonian respectively. The former encodes the optimization problem to be solved and the latter is required in case the initial state is an eigenstate of \(H_C\). In the absence of a mixing Hamiltonian, the algorithm might get stuck in one of the eigenstates of \(H_C\).
The optimal parameters \(\gamma_1^*,\beta_1^*,....\gamma_p^*,\beta_p^*\) are found using a classical optimization scheme so as to minimize the cost function.

\subsection{CVaR VQE and QAOA}
Conditional Value at Risk(CVaR), frequently used in finance, is a measure of risk associated with an investment. More precisely it quantifies the expected loss associated with a certain level of risk. It was shown that using CVaR expectation value as the loss function leads to a greater chance of getting a state that overlaps with the ground state when used with VQE and QAOA algorithms.\cite{c9}
CVaR expectation value is calculated by averaging the energies of the lowest \(\alpha\) \% of sampled energies. Suppose, the energies \(E_i\) are arranged in ascending order then CVaR expectation value is given by,
\begin{equation}
    CVaR=\frac{1}{\frac{\alpha N}{100}}\sum_{i=0} ^{i=\alpha N} E_i
\end{equation}
where \(n\) is the number of shots i.e. the number of energies measured.
%\subsection{CVaR QAOA}
%\subsection{Adapt VQE}

\subsection{Significance of protein selected}
The amino acid sequence chosen "YPYFIP" is a subsequence of the peptide chain that binds with the exosite of BACE1 Alzheimer's enzyme \cite{c10}. BACE1 has been shown to produce beta-amyloid which is responsible for symptoms like memory loss in Alzheimer's. We have removed the Acetylene and NH\textsubscript{2} chains and only considered the main peptide.
\subsection{Alphafold}
Alphafold is a state of the art deep learning model developed by Deepmind that predicts the 3-D structure of a protein based on the amino acid sequence. The model uses biological, evolutionary and geometric constraint of protein structures. This approch is quite different from the one used by Qiskit nature, the subject of this study where the approach is to minimize the interaction energy.
AlphaFold completely outperformed other techniques in CASP14, an annual competition that is considered the gold standard for protein folding \cite{c1}.
%%%%%%%%
%%%%%%%%
\section{Methodology and Setup} \label{meth_set}
\subsection{Hardware} 
We run our experiments on quantum devices provided by \textbf{IBM} and \textbf{Oxford Quantum Computing(OQC)}. Further, we tried using Rigetti and IONQ hardware but ran into compatibility issues and thus had to abandon them. The IBM, Rigetti, and OQC devices are based on superconducting qubits, while the IONQ device is based on Trapped Ion technology \cite{c11}. Another notable feature of IONQ QPU is that it has a fully connected topology thus any two-qubit gate can be directly applied to any qubit pair.

We alternatively also run the experiments on simulators to compare and contrast the results obtained on real hardware. We use the \textbf{SV1 simulator} provided by AWS braket, and \textbf{IBM FakeBrooklyn} provided by Qiskit. SV1 is a state vector simulator while FakeBrooklyn is a mock hardware with noise.

\begin{figure}
\centering
\begin{tabular}{|c|c|}
\hline
\subfloat{OQC LUCY}&
\subfloat{\centering \includegraphics[width = 2in]{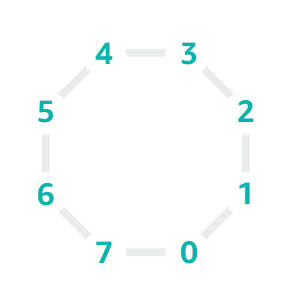}} \\
\hline
\subfloat{IBM Perth}&
\subfloat{\includegraphics[width = 2in]{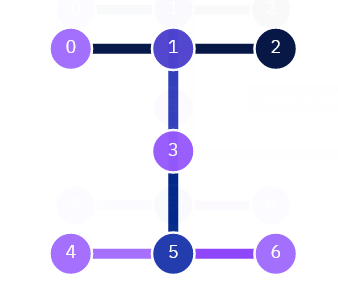}}
\\
\hline
\end{tabular}
\caption{Hardware Architecture of OQC and IBM Perth}
\end{figure}

\subsection{Ansatz}
For VQE we use Real Amplitudes Ansatz with a single repetition provided by Qiskit. Real Amplitudes ansatz has alternate layers of RY gates and CX gates. Default full entanglement map was used which applies the CX gate to all the qubit pairs.

For QAOA and CVaR QAOA, the ansatz is as discussed in \ref{lit_rev}. We experiment with different numbers of repetitions. Theoretically, as the number of repetitions increases, the solution should converge to the exact solution. However, it was found that the best results were obtained for p=2 repetitions. For all cases except OQC Lucy(see Section \ref{discussion}) p=2 was chosen.
\begin{figure}
    \centering
    \includegraphics[width=3in]{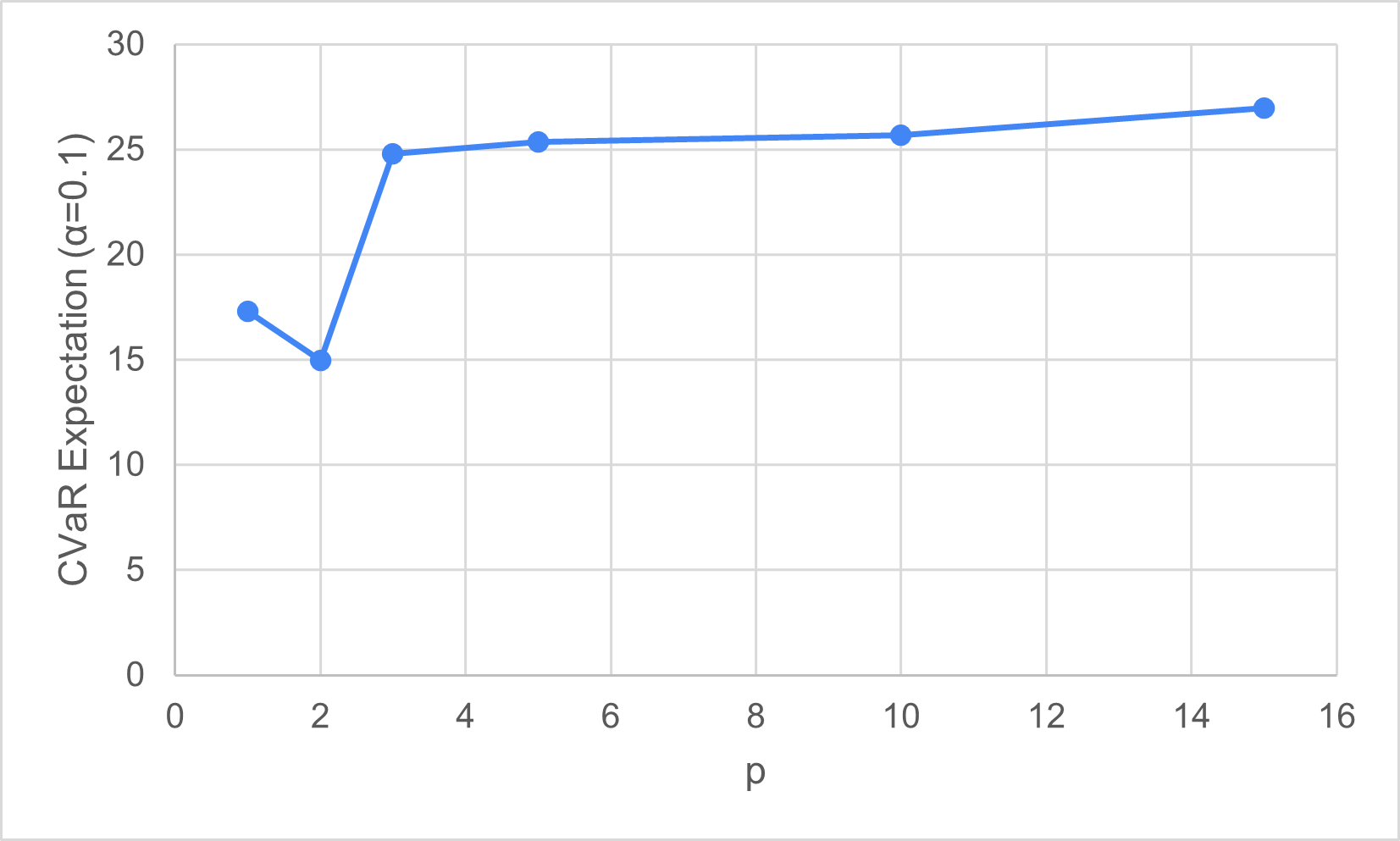}
    \caption{QAOA performance vs Ansatz repetitions. p represents the depth of the QAOA circuit.}
    \label{fig:qaoa_p}
\end{figure}

\subsection{Optimizer}
We use the COBYLA optimizer with 50 iterations. COBYLA is a classical gradient-free optimizer and calls the cost function only once per iteration. This is important from a cost point of view. Other optimizers such as gradient descent or SPSA could have been possibly chosen; however, in the interest of uniformity across different experiments and relatively erratic behaviour of other optimizers COBYLA was chosen.
\subsection{Error Mitigation}
NISQ devices are typically noisy and there exist several methods to mitigate/correct for these effects. There are two current approaches, the first being error correction, which involves methods to remove error during computation. On the other hand the error mitigation methods try to compensate for the errors post measurement.
We have used the error mitigation technique provided by qiskit simulating experiments on IBM Perth and FakeBrooklyn - noisy backend. A calibration matrix is calculated by fitting the measured state and the idea state data. The calibration matrix is then used to correct the states used in optimization.
%%%%%%%%%%%%%%%%%%%%%%%%%%%%%
%%%%%%%%%%%%%%%%%%%%%%%%%
\section{Results} \label{result}
\subsection{Ground State Energies on Hardware and Simulators}
The tables \ref{table:1} through \ref{table:5} summarize the ground state energies(in Hartrees) as achieved on different quantum hardware and simulators across the four algorithm types. Typically CVaR Expectation by definition is less than the expectation value. \(\alpha\)=0.01 was set for all cases except for IBM FakeBrooklyn, where \(\alpha\)=0.1 was chosen to compensate for fewer shots. The reduced number of shots for IBM FakeBrooklyn stems from memory and run-time constraints. 

CVaR VQE and CVaR QAOA seem to converge to the same value accross hardwares and simulators for simulations without the side chain as can be seen in Tables 1, II, and III. Error mitigation has no effect in this case.
For QAOA and VQE significant variations were noted. VQE in general performed better than QAOA, with VQE showing improvements when error mitigation was applied.

For simulations with the sidechain, VQE and CVaR VQE seem to outperform their QAOA counterparts as is evident in Tables IV and V.

%%%%%%%%%%%%%%%%%%%%%%%%%%%%%%%%%%%%%%%%%%%%%%%
\begin{table*}[htbp]
\begin{center}
\setlength{\arrayrulewidth}{0.25mm}
\setlength{\tabcolsep}{20pt}
\renewcommand{\arraystretch}{1.5}
\begin{tabular}{|c|c|c|}
 \hline
 \multicolumn{3}{|c|}{IBM PERTH} \\
 \hline
Algorithm & \vtop{\hbox{\strut Ground State}\hbox{\strut Energy}}& \vtop{\hbox{\strut Error Mitigated}\hbox{\strut Energy}}\\
 \hline
 VQE  &391.166&86.89\\
 \hline
CVaR VQE&-1.019&-1.019\\
 \hline
 QAOA &986.73&832.81\\
 \hline
 CVaR QAOA &-1.019&-1.019\\
 \hline
\end{tabular}

\end{center}
\caption{Table 1 shows the results of experiments ran on IBM Perth Hardware. The main chain is YPYFIP and there is no side chain used for this experiment. It took 6 qubits and the number of shots used are 8192.}
\label{table:1}
\end{table*}
%%%%%%%%%%%%%%%%%%%%%%%%%%%%%%%%%%%%%%%%%%%%%%%%

%%%%%%%%%%%%%%%%%%%%%%%%%%%%%%%%%%%%%%%%%%%%%%%%%%%%%
%%%%%%%%%%%%%%%%%%%%%%%%%%%%%%%%%%%%%%%%%%%%%%%%%%%%%%%%%%%%%%

%%%%%%%%%%%%%%%%%%%%%%%%%%%%%%%%%%%%%%%%%%%%%%%%%%%%%%%%%%
% \\[16pt]
\begin{table*}[htbp]
\begin{center}
\setlength{\arrayrulewidth}{0.25mm}
\setlength{\tabcolsep}{20pt}
\renewcommand{\arraystretch}{1.5}
\begin{tabular}{|c|c|c|}
 \hline
\multicolumn{3}{|c|}{Fake Brooklyn} \\
 \hline
Algorithm & \vtop{\hbox{\strut Ground State}\hbox{\strut Energy}}& \vtop{\hbox{\strut Error Mitigated}\hbox{\strut Energy}}\\
 \hline
 VQE  &   114.91  &84.22\\
 \hline
CVaR VQE& -1.019  &-1.019\\
 \hline
 QAOA &623.62  &666.46\\
 \hline
 CVaR QAOA &-1.019 &-1.019\\
 \hline
\end{tabular}

\end{center}
\caption{Table 2 shows the results of experiments ran on Fake Brooklyn Simulator. The main chain is YPYFIP and there is no side chain used. It took 6 qubits and the number of shots used are 128. }
\label{table:3}
\end{table*}
%%%%%%%%%%%%%%%%%%%%%%%%%%%%%%%%%%%%%%%%%%%%%%%%%%%%%%%%%%%%%%%%%%%%%%%%%%%

%%%%%%%%%%%%%%%%%%%%%%%%%%%%%%%%%%%%%%%%%%%%%%%%%%%%%%%%%%%%%%%%%%%%%%%%%%%%%%%%%%
\begin{table*}[htbp]
\begin{center}
\setlength{\arrayrulewidth}{0.25mm}
\setlength{\tabcolsep}{20pt}
\renewcommand{\arraystretch}{1.5}
\begin{tabular}{|c|c|}
 \hline
\multicolumn{2}{|c|}{Lucy} \\
 \hline
Algorithm & \vtop{\hbox{\strut Ground State}\hbox{\strut Energy}} \\
 \hline
 VQE  &   768.7303  \\
 \hline
CVaR VQE&   -1.019\\
 \hline
 QAOA &746.8159 \\ %Change
 \hline
 CVaR QAOA &-1.019 \\ %Change
 \hline
\end{tabular}

\end{center}
\caption{Table 3 shows the results of experiments ran on Lucy. There is no error mitigation technique used in running this experiment. The main chain is YPYFIP and the there is no side chain used. It took 6 qubits and the number of shots used are 8192. }
\label{table:4}
\end{table*}
%%%%%%%%%%%%%%%%%%%%%%%%%%%%%%%%%%%%%%%%%%%%%%%%%%%%%%%%%%%%%%%%%%%%%%%
% \\[16pt]
\begin{table*}[htbp]

\begin{center}
\setlength{\arrayrulewidth}{0.25mm}
\setlength{\tabcolsep}{20pt}
\renewcommand{\arraystretch}{1.5}
\begin{tabular}{|c|c|}
 \hline
\multicolumn{2}{|c|}{Fake Brooklyn} \\
 \hline
Algorithm & \vtop{\hbox{\strut Ground State}\hbox{\strut Energy}}\\
 \hline
 VQE  &1601.5051 \\
 \hline
CVaR VQE&17.1838\\
 \hline
 QAOA &2123.2193\\
 \hline
 CVaR QAOA &27.562968\\
 \hline
\end{tabular}

\end{center}
\caption{Table 4 shows the results of experiments run on Fake Brooklyn Simulator. There is no error mitigation technique used in running this experiment. The main chain is YPYFIP and the side chain is IPFY. It took 17 qubits and the number of shots used are 128. }
\label{table:2}
\end{table*}

%%%%%%%%%%%%%%%%%%%%%%%%%%%%%%%%%%%%%%%%%%%%%%%%%%%%%%%%%%%%%%%%%%%%%%%%
%qiskit_cvar_qaoa_0_01_sv1_50_iter_depth_2_peptide
% \\[16pt]
\begin{table*}[htbp]
\begin{center}
\setlength{\arrayrulewidth}{0.25mm}
\setlength{\tabcolsep}{20pt}
\renewcommand{\arraystretch}{1.5}
\begin{tabular}{|c|c|}
 \hline
 \multicolumn{2}{|c|}{SV1 Simulator} \\
 \hline
Algorithm & \vtop{\hbox{\strut Ground State}\hbox{\strut Energy}}\\
 \hline
 VQE  &  226.27  \\
 \hline
CVaR VQE&   -1.019  \\
 \hline
 QAOA &1741.854  \\
 \hline
 CVaR QAOA &8.33775  \\
 \hline
\end{tabular}
\end{center}
\caption{Table 5 shows the results of experiments run on SV1 Simulator. There is no error mitigation technique used in running this experiment. The main chain is YPYFIP, and the side chain is IPFY. It took 17 qubits and the number of shots used were 8192. }
\label{table:5}
\end{table*}
% \\[16pt]

\begin{figure}
     \centering
     \includegraphics[width=3.4in]{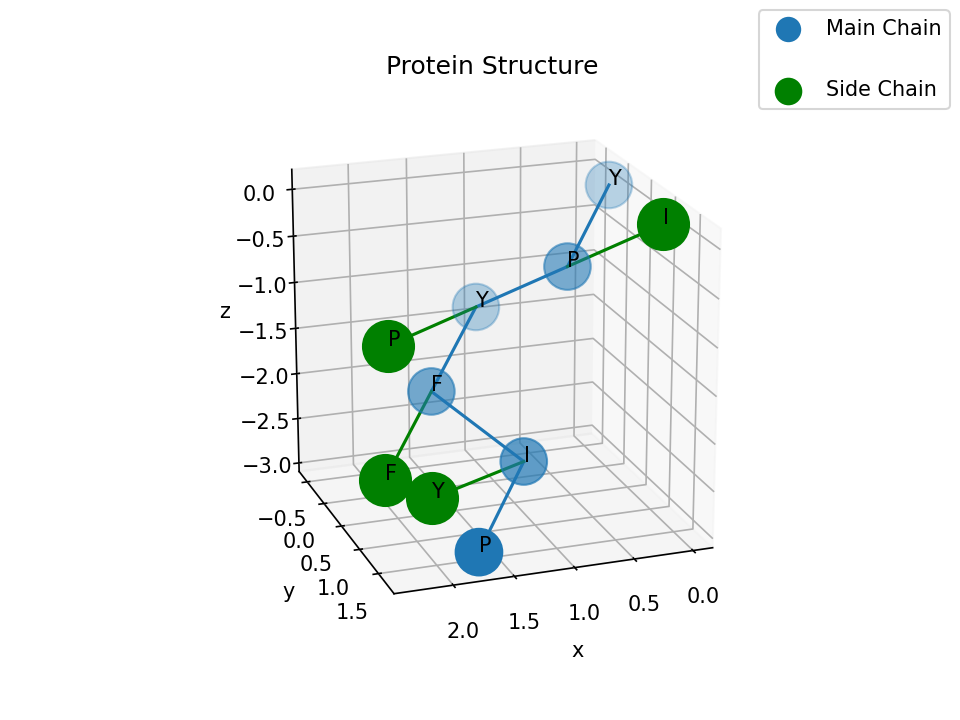}
     \caption{Sample protein structure with sidechain }
     \label{fig:my_label}
\end{figure}

\subsection{Distribution of Algorithm Predictions}
We study the repeatability of each of the four algorithms. Thus to quantify the variation in results, we ran experiments on a noiseless simulator with the protein chain without any side chains. We characterize the distribution of predicted ground state energy(or CVaR expectation in the case of CVaR variants) and the predicted protein structure across multiple trials to quantify the efficacy of the algorithms.
The predicted structures show considerable variability as is evident in the figures \ref{fig:1} \ref{fig:2} \ref{fig:3} \ref{fig:4}.  
The results thus, not only help in quantifying the efficacy of the method in its ultimate goal which is to predict the 3-D structure but also provide a more robust way of comparing the predicted structures, as will be discussed in the next section.
VQE and CVaR QAOA were statistically the most consistent in terms of the returning the same structure. However, both of them predicted physically impossible structures for a significant number of trials.
\begin{figure}
    \begin{tabular}{cc}
    \subfloat[Distribution of predicted eignevalue]{\includegraphics[width = 1.5in]{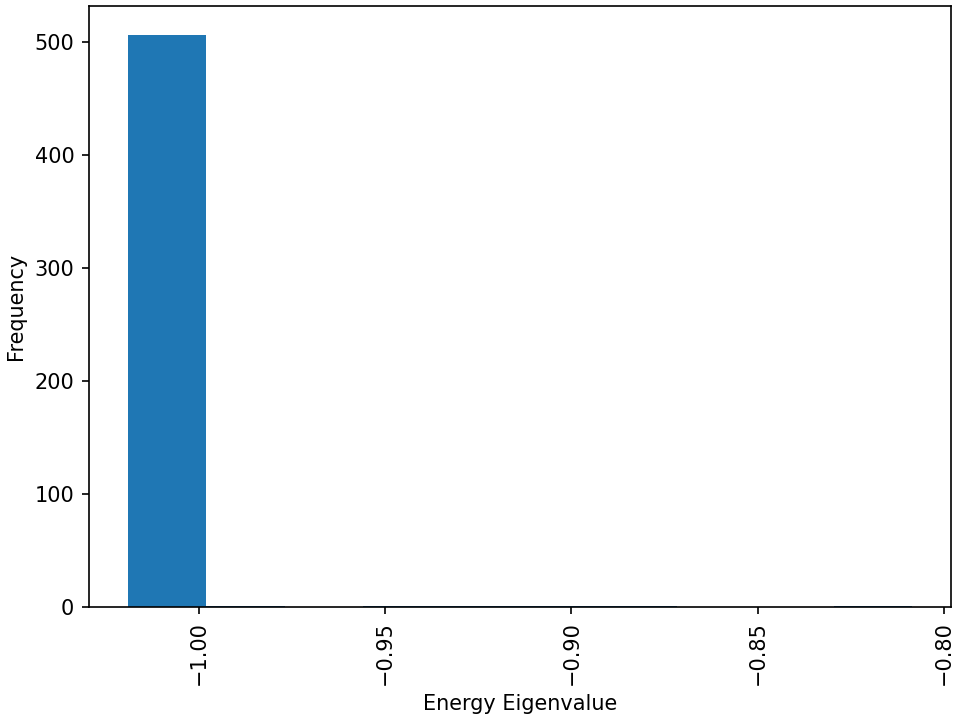}} &
    \subfloat[Distribution of predicted peptide folding, the x-axis labels denote the turns as defined in \ref{lit_rev} ]{\includegraphics[width = 1.5in]{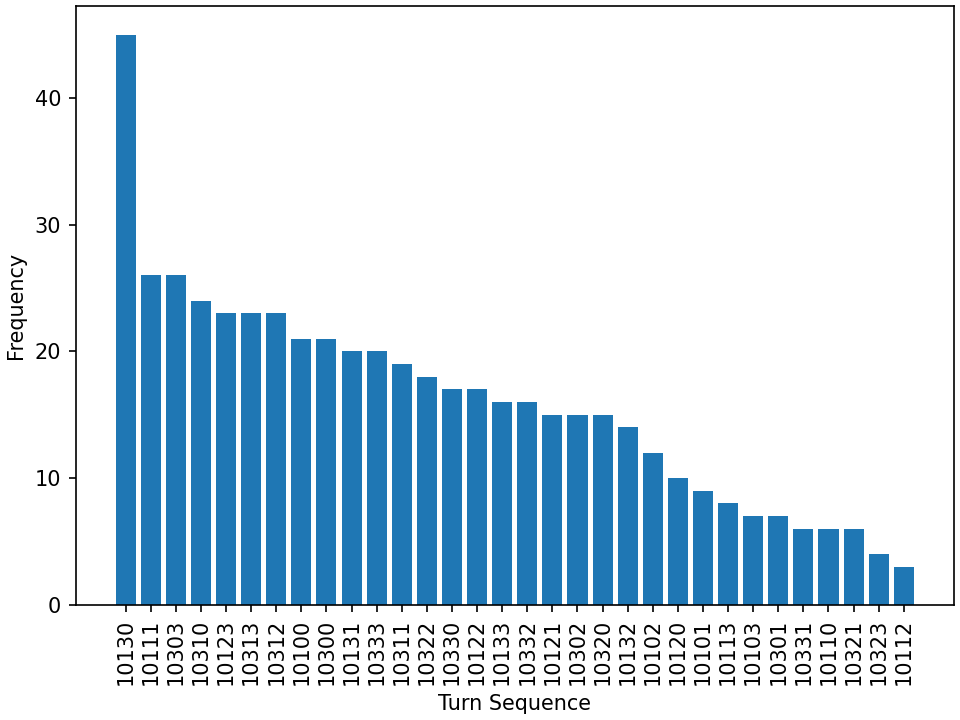}} \\
    \subfloat[Corresponding predicted structure]{\includegraphics[width = 1.5in]{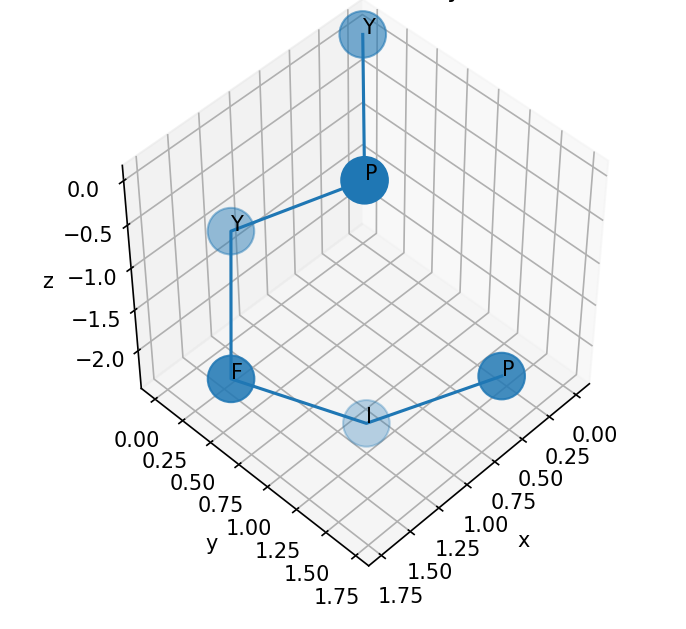}} &
    \subfloat[Second ranked prediction]{\includegraphics[width=1.5in]{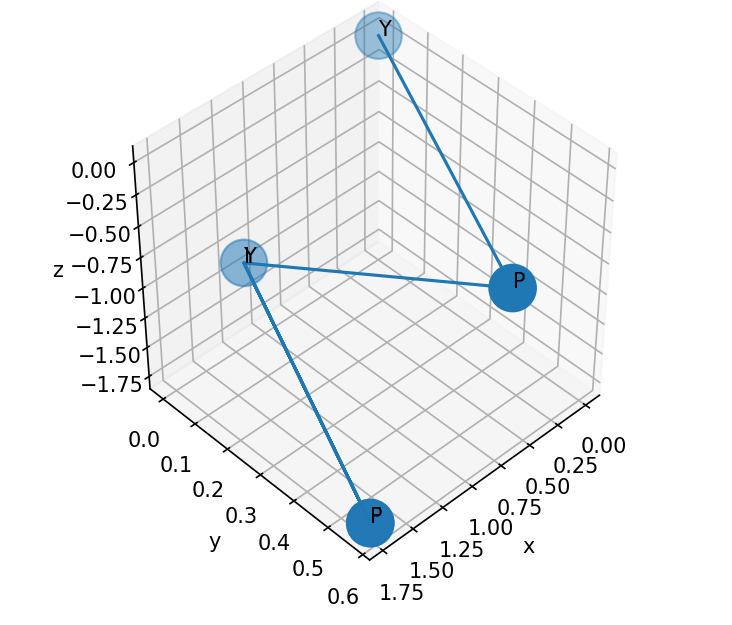}}

    \end{tabular}
    \caption{CVaR QAOA: The histogram(a) summarises the predicted results as obtained on the QASM simulator for CVaR QAOA (\(\alpha\)=0.01). The prediction was made 512 times; the eigenvalue converged to the same value 98.8\% of the time. However, the most frequent prediction had only a frequency of 8.8\%. Moreover, Figure (d) indicates that the chain folds back onto itself leading to overlap, something that is forbidden.}
    \label{fig:1}
\end{figure}

\begin{figure}
    \begin{tabular}{cc}
    \subfloat[Distribution of predicted eigenvalue]{\includegraphics[width = 1.5in]{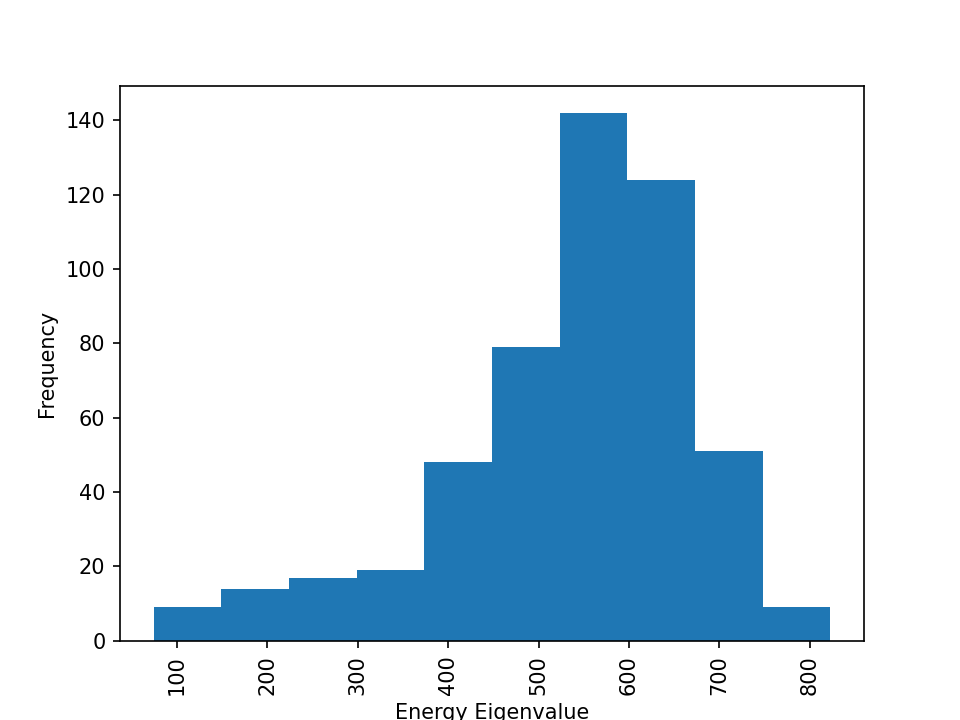}} &
    \subfloat[Distribution of predicted protein folding, the x-axis labels denote the turns as defined in \ref{lit_rev}]{\includegraphics[width = 1.5in]{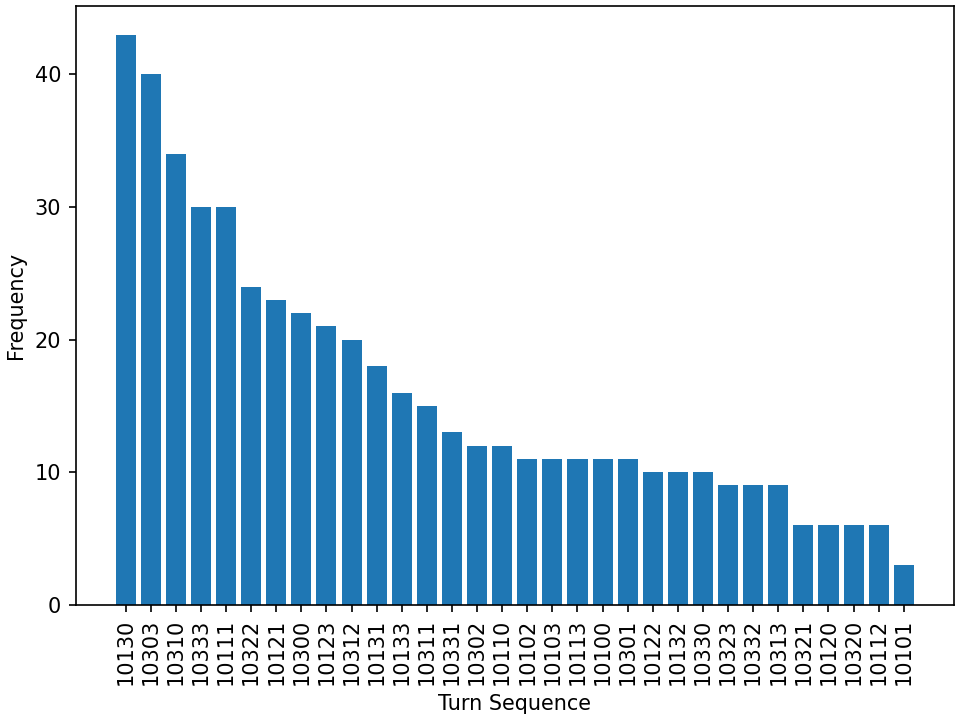}} \\
    \subfloat[First ranked prediction]{\includegraphics[width=1.5in]{Figures/aer_130_qaoa.PNG}}&
    \subfloat[Second ranked prediction]{\includegraphics[width=1.5in]{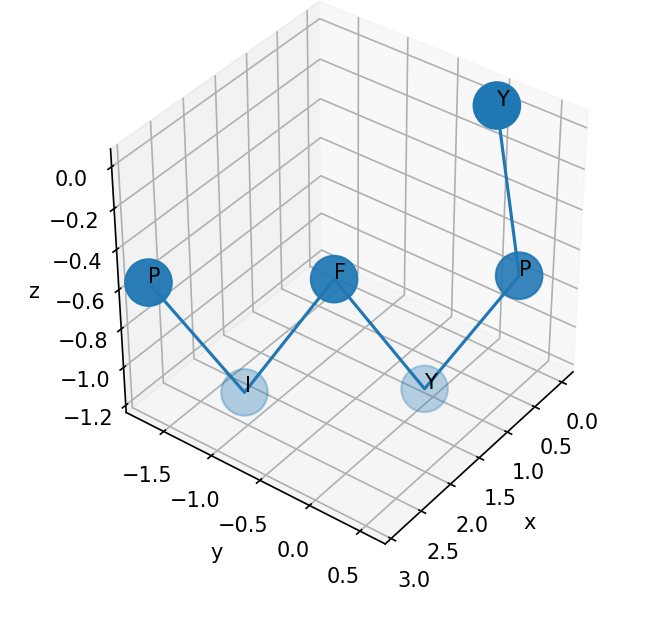}}

    \end{tabular}
    \caption{QAOA: The histogram(a) summarises the predicted results as obtained on the QASM simulator for QAOA. The prediction was made 512 times, the algorithm converged to the lowest energy state less than 2\% of the time. The most frequent prediction had only a probability of 8.4\%. The top two most frequent predictions do not show signs of overlap.}
    \label{fig:2}
\end{figure}

\begin{figure}
    \begin{tabular}{cc}
    \subfloat[Distribution of predicted eignevalue]{\includegraphics[width = 1.5in]{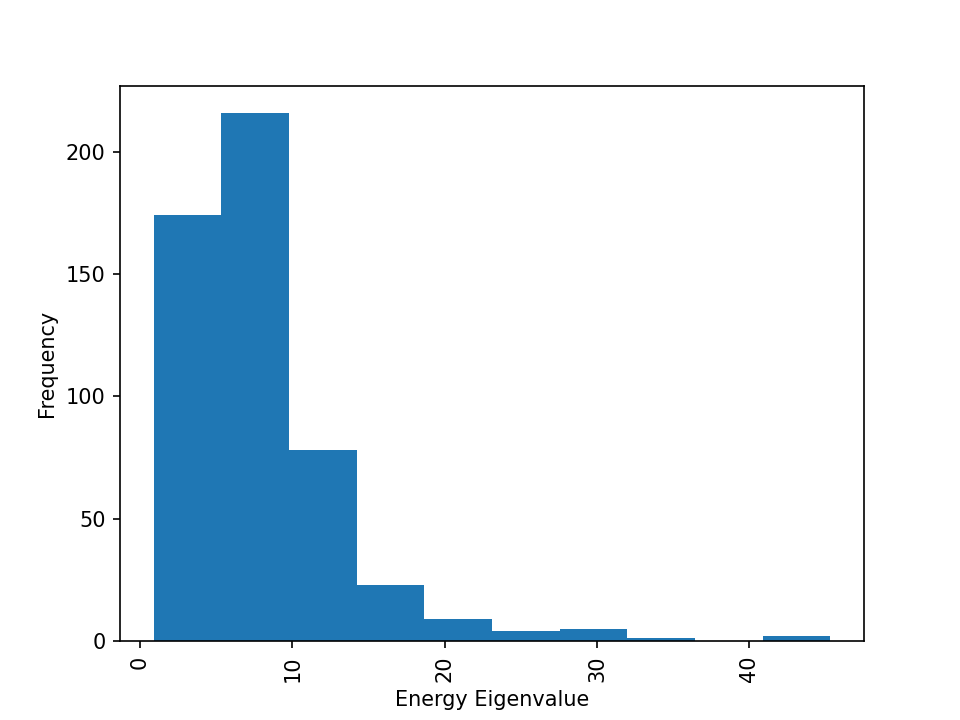}} &
    \subfloat[Distribution of predicted protein folding, the x-axis labels denote the turns as defined in \ref{lit_rev}]{\includegraphics[width = 1.5in]{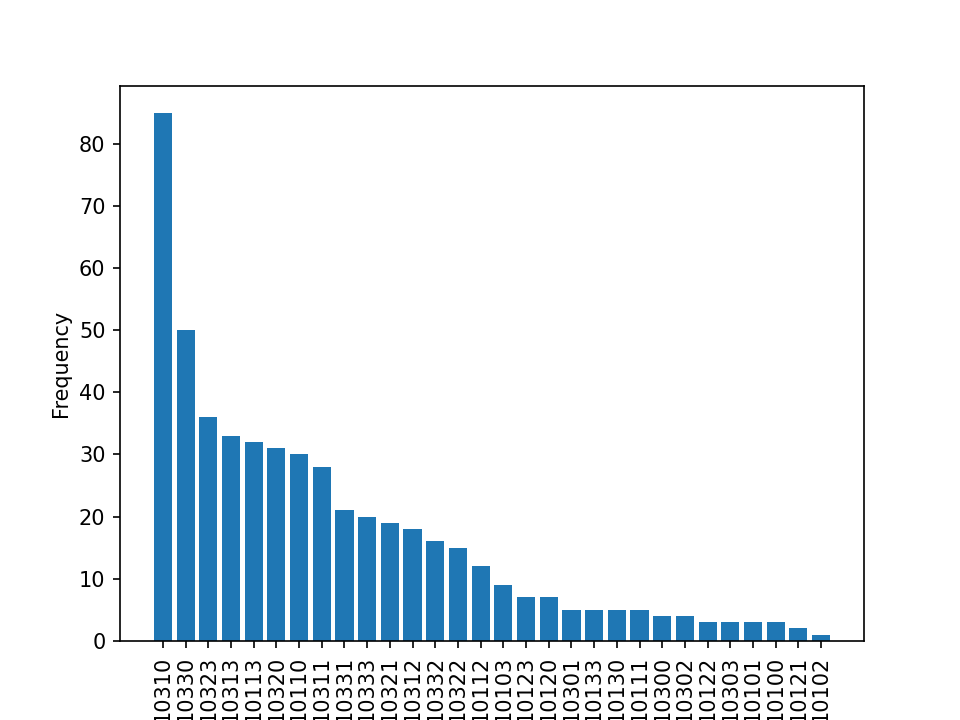}} \\
    \subfloat[First ranked prediction]{\includegraphics[width=1.5in]{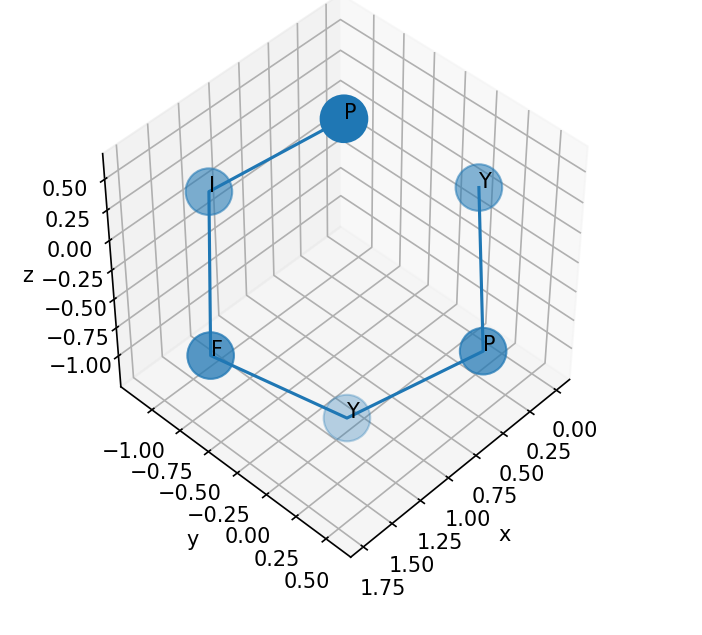}}&
    \subfloat[Second ranked prediction]{\includegraphics[width=1.5in]{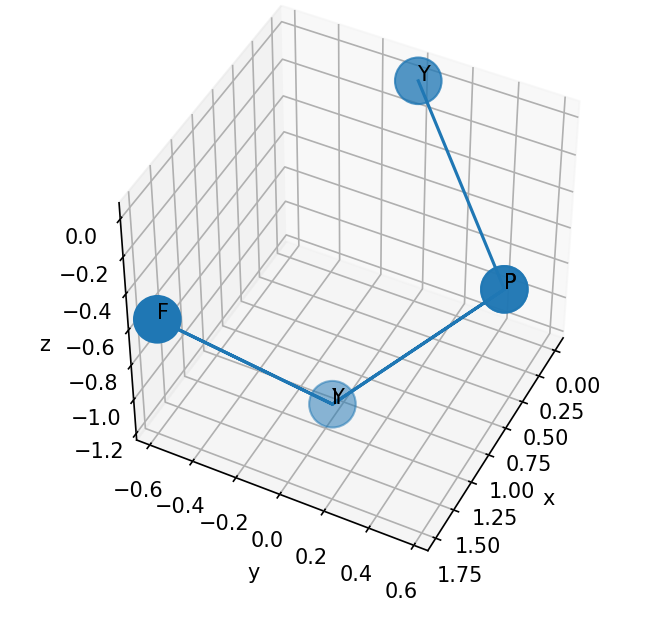}}

    \end{tabular}
    \caption{VQE: The histogram(a) summarises the predicted results as obtained on the QASM simulator for VQE. The prediction was made 512 times, the algorithm converged to the lowest energy state ~30\% of the time. The most frequent prediction had a frequency of 16\%. Despite promising numbers, Figure (d) indicates that the chain folds back onto itself leading to overlap, something that is forbidden.}
    \label{fig:3}
\end{figure}

\begin{figure}
    \begin{tabular}{cc}
    \subfloat[Distribution of predicted eignevalue]{\includegraphics[width = 1.5in]{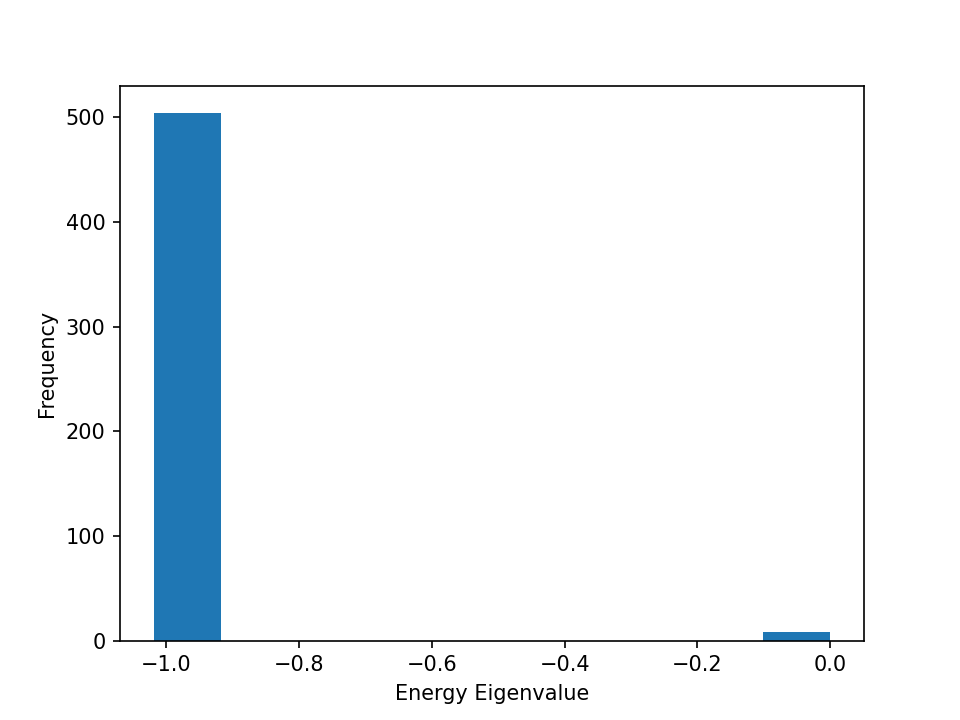}} &
    \subfloat[Distribution of predicted protein folding, the x-axis labels denote the turns as defined in \ref{lit_rev}]{\includegraphics[width = 1.5in]{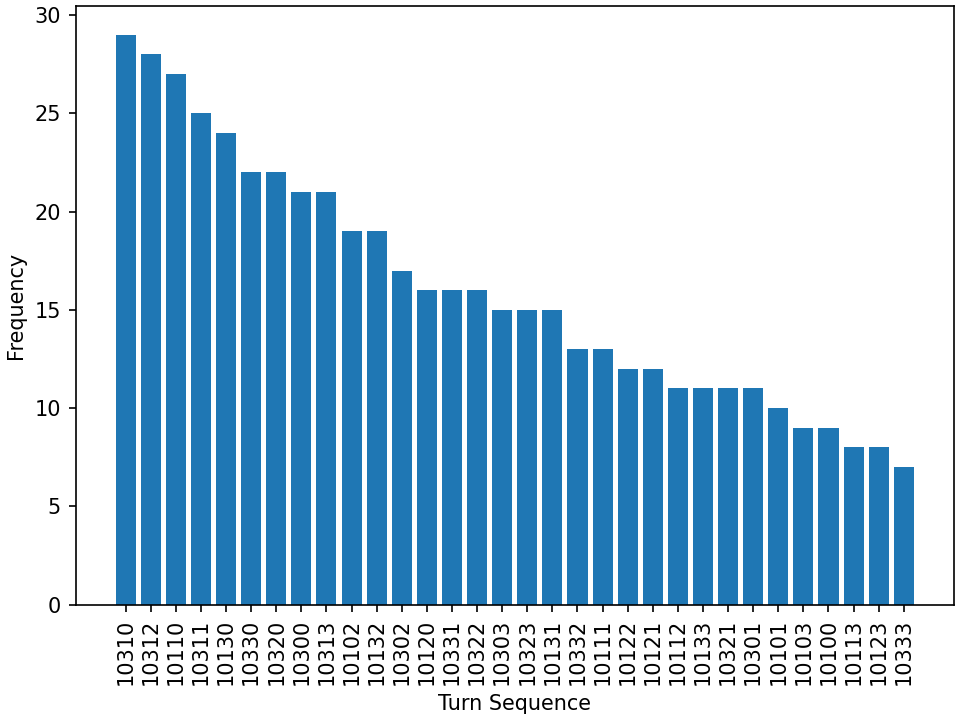}} \\
    \subfloat[First ranked prediction]{\includegraphics[width=1.5in]{Figures/aer_310_vqe.PNG}}&
    \subfloat[Second ranked prediction]{\includegraphics[width=1.5in]{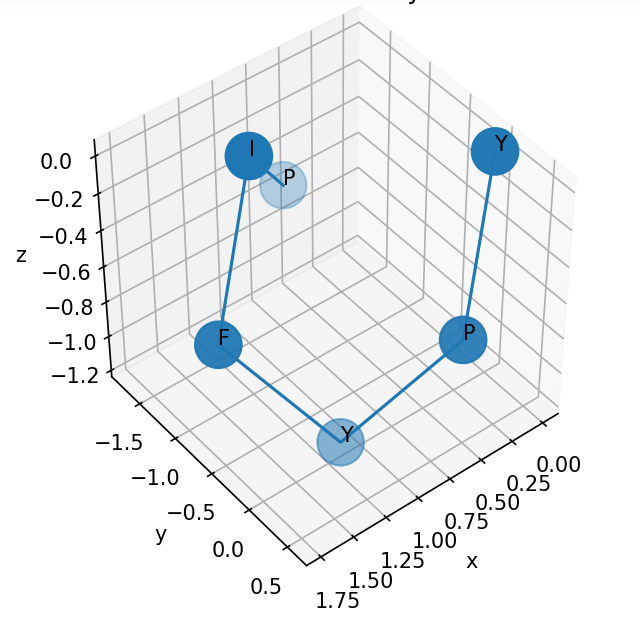}}

    \end{tabular}
    \caption{CVaR VQE: The histogram(a) summarises the predicted results as obtained on the QASM simulator for VQE. The prediction was made 512 times, the algorithm converged to the lowest energy state ~99\% of the time. The most frequent prediction had a frequency of only 5\%. Moreover, numbers, Figure (d) indicates that the chain folds back onto itself leading to overlap, something that is forbidden.}
    \label{fig:4}
\end{figure}

\subsection{RMSD Calculation}
Root Mean Square Deviation(RMSD) is the most commonly used distance-based measurement tool to compare the protein structure similarity between two atomic coordinates. RMSD values are calculated by the formula \\
\begin{equation}
    \sqrt{1/n\sum_{i=1}^{n} d_i^{2}}\
\end{equation}

It is to be noted that the any RMSD value less that 2 {\AA} is a good value, nevertheless, the lesser the value, the higher the accuracy. We compare the top predictions by QAOA class of algorithms and VQE class of algorithms by computing the RMSD \cite{c12}. The RMSD value turned out to be 0.6357{\AA} between the QAOA class and VQE class (Fig \ref{QAOA vs VQE}).

\begin{figure}
    \begin{tabular}{cc} 
    \subfloat[Protein Structure with QAOA]{\includegraphics[width = 1.5in]{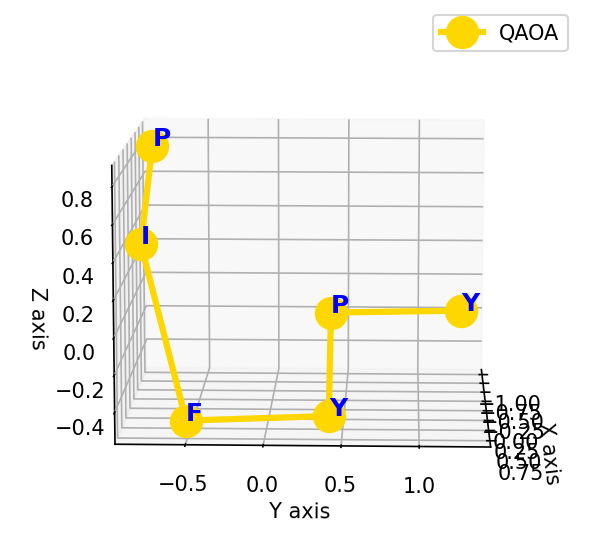}}&
    \subfloat[Protein Structure with VQE]{\includegraphics[width=1.5in]{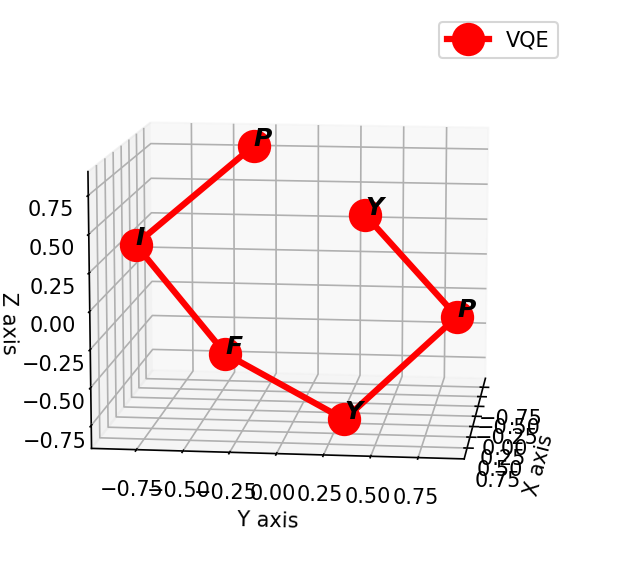}}\\
    \subfloat[Comparing the structures of QAOA and VQE]{\includegraphics[width=1.5in]{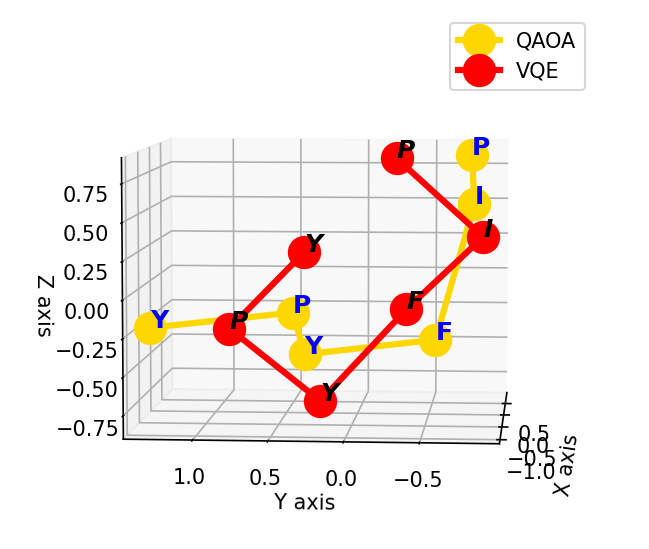}}&
    \subfloat[Comparing AlphaFold Predictions(Red) with VQE top prediction(Yellow)]{\includegraphics[width=1.5in]{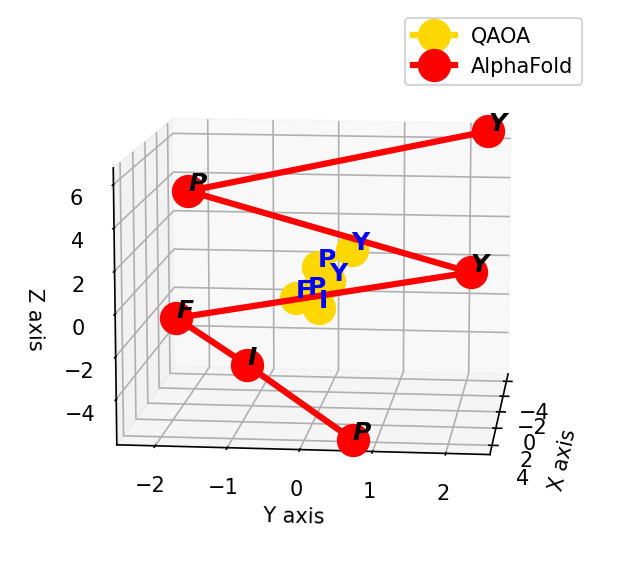}}

    \end{tabular}
    \caption{Comparing most probable structures predicted by QAOA and VQE}
    \label{QAOA vs VQE}
\end{figure}
%Qiskit Nature with
Further we compare the predictions made by AlphaFold \cite{c1} with our top predictions using the same method. In order to make the two predictions comparable, alpha fold prediction (Figure \ref{fig:alphafold}) was reduced to a bead like model by considering the peptide as a chain of geometric centers of each amino acid. The comparison is illustrated in Figure \ref{QAOA vs VQE} (d). RMSD value of 4.795{\AA} was obtained.

There is a noticeable difference in the scale of the two images. We conjecture this is because of the lattice parameter inbuilt in Qiskit Nature. 
\subsection{Effect of penalty terms}
The effects of the three penalty terms described in Section \ref{lit_rev} are studied. Each of the three terms is varied on a logarithmic scale keeping other terms constant and the probability of the most frequent structure is noted (Fig \ref{fig:my_label}). 
\begin{figure}
    \centering
    \includegraphics[width=3in]{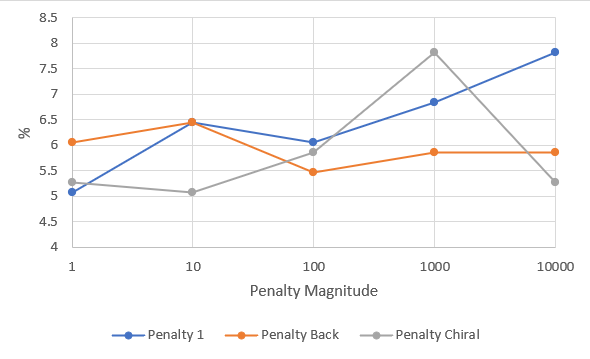}
    \caption{\% probability of the most frequent structure plotted against the penalty term magnitude as defined in Section \ref{lit_rev}}
    \label{fig:my_label}
\end{figure}

\section{Discussion} \label{discussion}
The cost of executing the VQE algorithm scales with two factors; the number of quantum tasks and the number of shots. The former being the costlier of the two. These two correspond to the number of circuits created and the number of shots executed on each circuit. If the Hamiltonian terms commute it is possible to measure them simultaneously thus reducing the number of calls made to the quantum computer, and reducing the major cost component. However, if the terms do not commute, one has to measure each hamiltonian term separately which scales as \(O(N^4)\). In our case, all Hamiltonian terms were composed of Pauli Z matrices. Our cost thus only scaled with the number of shots. 

Another important factor that determines the capability of the hardware to simulate is the pulse duration. OQC Lucy allows a maximum pulse duration for 120 \(\mu\)s thus limiting the circuit depth to 1 at 8192 shots for QAOA.

The number of iterations is also important to consider. Optimizers such as gradient descent or SPSA might make more than one call to the quantum computer per iteration further increasing the cost multifold(typically three-fold for gradient-based optimizers as they require the calculation of gradient by parameter shift.)

To reduce the computational costs, the proteins qubit operator has been reduced to the minimum amount of qubits required by omitting identity operations. For the protein with the main chain, that has been used in the paper, the problem is mapped to 120 qubits but it has been optimized to 17 qubits. Similarly, it has been optimized to 6 qubits in the case without a chain. 

The structures obtained using the Variational Algorithms differ considerably from similar techniques. To illustrate, we include the structures predicted by alpha fold and alpha fold 2.
\begin{figure}
    \centering
    \includegraphics[width = 3in]{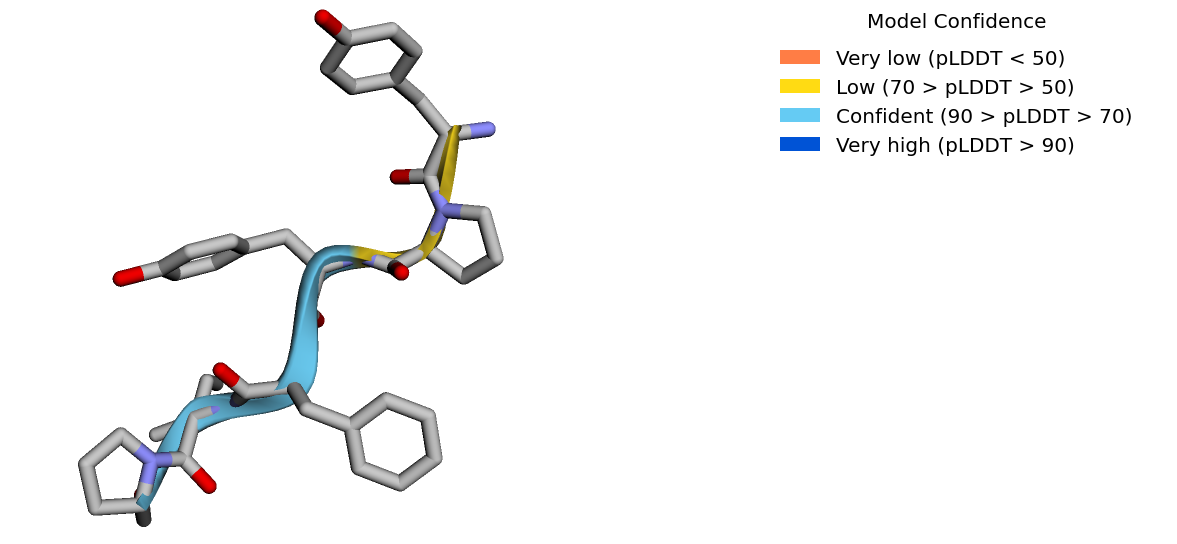}
    \caption{Structure Predicted by AlphaFold}
    \label{fig:alphafold}
\end{figure}

\begin{figure}[!t]
    \centering
    \includegraphics[width=3in]{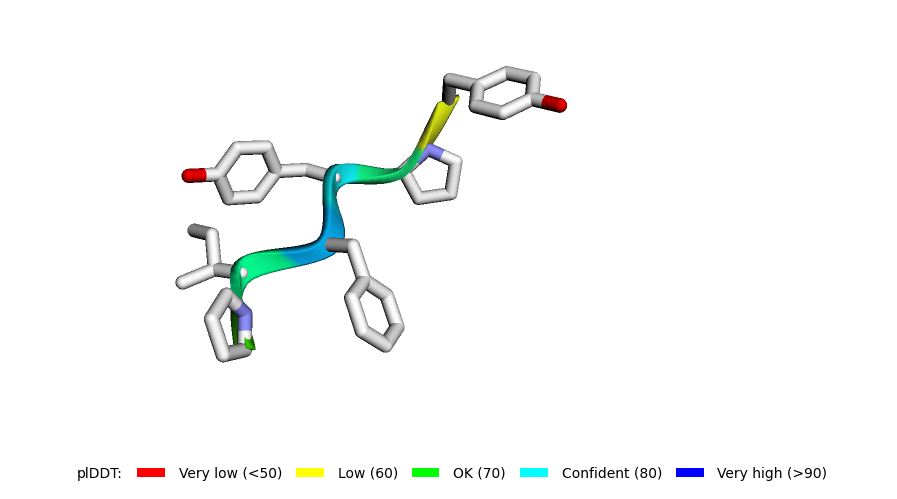}
    \caption{Structure Predicted by AlphaFold 2}
    \label{fig:alphafold2}
\end{figure}

The reliability of the model is quite low, a probable cause might be that the energy difference between different structures is not substantial. Further, since only a limited number of interactions were considered, this leads to inaccuracies. 
A pressing problem with the current model is that chain folds onto itself on multiple occasions.
\section{Conclusion} \label{conclusion}
In the paper, we analyzed the performance of the protein folding model provided by Qiskit as a part of the Qiskit Nature library and compare the results against the Alphafold model and find that there are significant deviations which require that the Qiskit Nature model be improved. The algorithm is one of a kind, especially in the paradigm of gate-based quantum computing.

We compare VQE and QAOA on various hardware and simulators, with and without error mitigation. We conclude that VQE class of algorithms performed better than QAOA for simulations with and without a side chain. We also compare the structures that were predicted by the algorithms and go on to examine the repeatability of the model we find that several unphysical predictions were being made frequently. Further, we also examined the effect of penalty parameters on the statistical stability of the model and find a weak trend in the same. 

We conclude that there is an extensive scope of improvement on various metrics, however, we also acknowledge that the current limitations on the hardware make it difficult to achieve. Continuous work in this direction is a positive indication. A more robust encoding method that prevents folding of the chain back onto itself must be developed among other things. 
\section{Future Research} \label{future_research}
Future developments in this field would closely follow the availability of hardware capabilities. The primary goal for future works is expected to be modelling the problem in a way that allows more degrees of freedom and prohibits unphysical states. Further, the reliability of the model must be improved to allow for commercial and practical use.

One avenue that we identify is the use of other emerging techniques such as Density Matrix Embedding Theory to better characterize the energies of the proteins.
%dmet
\section*{Acknowledgments}
%We are grateful to Fractal Analytics for supporting this research. 
The authors acknowledge using Google Colab Pro and AWS Braket for carrying out the experiments and the libraries Pennylane and Qiskit Nature from IBM Quantum.  
%Bibliography
\bibliographystyle{IEEEtran}  
% \bibliography{IEEEabrv,References}  

\begin{thebibliography}{9}

\bibitem{c1} J. Jumper, R. Evans, A. Pritzel, T. Green, M. Figurnov, O. Ronneberger, K. Tunyasuvunakool, R. Bates, A. Potapenko, A. Bridgland, C. Meyer, S. A. A. Kohl, A. J. Ballard, A. Cowie, B. Romera- Paredes, S. Nikolov, R. Jain, J. Adler, T. Back, S. Petersen, D. Reiman, E. Clancy, M. Zielinski, M. Steinegger, M. Pacholska, T. Berghammer, S. Bodenstein, D. Silver, O. Vinyals, A. W. Senior, K. Kavukcuoglu, P. Kohli, and D. Hassabis, "Highly accurate protein structure prediction with AlphaFold," Nature, vol. 596, no. 7873, pp. 583–589, 2021.

\bibitem{c2} A. Robert, P. K. Barkoutsos, S. Woerner, and I. Tavernelli, “Resource-efficient quantum algorithm for protein folding,” npj Quantum Information, vol. 7, no. 1, feb 2021. [Online]. Available: https://doi.org/10.1038%2Fs41534-021-00368-4  

\bibitem{c3} K. A. Dill, S. B. Ozkan, M. S. Shell, and T. R. Weikl, “The protein folding problem,” Annual review of biophysics, vol. 37, p. 289, 2008.  

\bibitem{c4} M. S. R. K. J. E. Maria Kieferova, Paul K. Faehrmann, “Randomizing multi-product formulas for improved hamiltonian simulation,” arXiv preprint arXiv:2101.07808, 2021.  

\bibitem{c5} S. Miyazawa and R. L. Jernigan, “Residue–residue potentials with a favorable contact pair term and an unfavorable high packing density term, for simulation and threading,” Journal of molecular biology, vol. 256, no. 3, pp. 623–644, 1996.  

\bibitem{c6} P. K. Barkoutsos, G. Nannicini, A. Robert, I. Tavernelli, and S. Woerner, “Improving variational quantum optimization using CVaR,” Quantum, vol. 4, p. 256, apr 2020. [Online]. Available: https://doi.org/10.22331%2Fq-2020-04-20-256  

\bibitem{c7} A. Peruzzo, J. McClean, P. Shadbolt, M.-H. Yung, X.-Q. Zhou, P. J. Love, A. Aspuru-Guzik, and J. L. O’brien, “A variational eigenvalue solver on a photonic quantum processor,” Nature communications, vol. 5, no. 1, pp. 1–7, 2014.  

\bibitem{c8} E. Farhi, J. Goldstone, and S. Gutmann, “A quantum approximate optimization algorithm,” arXiv preprint arXiv:1411.4028, 2014.  

\bibitem{c9} P. K. Barkoutsos, G. Nannicini, A. Robert, I. Tavernelli, and S. Woerner, “Improving variational quantum optimization using cvar,” Quantum, vol. 4, p. 256, 2020.  

\bibitem{c10} L. J. Gutierrez, E. Angelina, A. Gyebrovszki, L. F, N. Peruchena, H. A. Baldoni, B. Penke, and R. D. Enriz, “New small-size peptides modulators of the exosite of bace1 obtained from a structure-based design,” Journal of Biomolecular Structure and Dynamics, vol. 35, no. 2, pp. 413–426, 2017.

\bibitem{c11} K. Wright, K. M. Beck, S. Debnath, J. Amini, Y. Nam, N. Grzesiak, J.- S. Chen, N. Pisenti, M. Chmielewski, C. Collins et al., “Benchmarking an 11-qubit quantum computer,” Nature communications, vol. 10, no. 1, pp. 1–6, 2019.  

\bibitem{c12} R. A. Irina Kufareva, “Methods of protein structure comparison,” Meth- ods in molecular biology (Clifton, N.J.), vol. 857, pp. 231–57, 2015. 


\end{thebibliography}

\end{document}